\documentclass[a4paper,10pt]{article} \input{epsf}

\usepackage[]{amssymb}
\usepackage{graphics}

\title{On the occurrence and detectability of\\
Bose-Einstein condensation\\ in helium white dwarfs}

\author{
O.~G. Benvenuto\footnote{Member of  the Carrera del Investigador Cient\'{\i}fico, 
Comisi\'on de  Investigaciones Cient\'{\i}ficas de la Provincia de Buenos Aires (CIC). 
Email: obenvenu@fcaglp.unlp.edu.ar} 
and M.~A.   De   Vito\footnote{Member of  the Carrera del Investigador Cient\'{\i}fico, 
Consejo Nacional de  Investigaciones Cient\'\i ficas y T\'ecnicas (CONICET). 
Email: adevito@fcaglp.unlp.edu.ar} 
\\
Facultad de Ciencias Astron\'omicas y Geof\'{\i}sicas\\ 
Universidad Nacional de La Plata\\ 
Paseo del Bosque S/N, B1900FWA, La Plata, Argentina\\ 
and \\
Instituto de Astrof\'{\i}sica de La Plata\\ 
IALP, CCT-CONICET-UNLP, Argentina
}

\begin{document}

\maketitle

\begin{abstract}

It has been recently proposed that helium white dwarfs may provide promising conditions for the occurrence of the Bose~-~Einstein condensation. The argument supporting this expectation is that in some conditions attained in the core of these objects, the typical De Broglie wavelength associated with helium nuclei is of the order of the mean distance between neighboring nuclei. In these conditions the system should depart from classical behavior showing quantum effects. As helium nuclei are bosons, they are expected to condense.

In order to explore the possibility of detecting the Bose~-~Einstein condensation in the evolution of helium white dwarfs we have computed a set of models for a variety of stellar masses and values of the condensation temperature. We do not perform a detailed treatment of the condensation process but mimic it by suppressing the nuclei contribution to the equation of state by applying an adequate function. As the cooling of white dwarfs depends on average properties of the whole stellar interior, this procedure should be suitable for exploring the departure of the cooling process from that predicted by the standard treatment.

We find that the Bose~-~Einstein condensation has noticeable, but {\it not} dramatic effects on the cooling process only for the most massive white dwarfs compatible with a helium dominated interior ($\approx 0.50 M_\odot$) and very low luminosities (say, $Log(L/L_\odot) < -4.0$). These facts lead us to conclude that it seems extremely difficult to find observable signals of the Bose~-~Einstein condensation. 

Recently, it has been suggested that the population of helium white dwarfs detected in the globular cluster NGC~6397 is a good candidate for detecting signals of the Bose~-~Einstein condensation. We find that these stars have masses too low and  are too bright to have an already condensed interior.

\end{abstract}

%------------------------------------------------------------------------------------------------------------------------
\section{Introduction}\label{sec:intro} 

It is well known that low and intermediate mass stars (up to $\approx 8 M_\odot$) evolve leading to the formation of white dwarf (hereafter WD) stars. Depending on their masses, WDs are expected to have different internal compositions. For example, low mass WDs should have a helium rich interior because they have not reached central temperatures high enough for the onset of helium burning  (T~$\approx10^8$~K). These objects have masses up to $\approx 0.5 M_\odot$ and are usually referred to as helium WDs (hereafter He~WDs). More massive WD ($0.5M_\odot \lesssim M \lesssim 1.1M_\odot$) are remnant of higher mass stars that underwent nuclear activity up to helium burning. Consequently, the interior  of these WDs is expected to be made up of a mixture of carbon and oxygen with traces of neon. Finally, the most massive WDs ($1.1M_\odot \lesssim M \lesssim 1.3M_\odot$) are expected to have suffered nuclear activity beyond helium burning, leading to the formation of an oxygen and magnesium rich interior. For a detailed discussion of the evolutionary processes leading to the formation of WDs see, e.g., Ref.~\cite{Iben}. In this paper we are mainly interested in He~WDs.

From a physical point of view \cite{shap-teuk,koes-chang}, WDs are remarkably simple objects. Their structure is supported against gravitational collapse by a Fermi gas of strongly degenerate electrons. Due to the high efficiency of degenerate material to conduct heat, cool WDs have nearly isothermal interiors. They evolve along an almost straight track in the Hertzsprung~-~Russell (luminosity vs. effective temperature) diagram. Because of its simplicity, many researchers studied the cooling of WDs with the aim of employing evolutionary calculations to measure, e.g., the age of the disk of our Galaxy \cite{winget87}. With the availability of very deep observations, thanks to the Hubble Space Telescope, it has been possible to study the WD population of globular clusters making WD evolution an important ingredient for the understanding of the evolution of all the stars in such clusters. These studies called for an increasing degree of sophistication in the modeling of WD cooling. It was recognized the relevance of element diffusion that modified the chemical composition of the outermost layers of the WD \cite{iben-macdonald85,asb2001}. This process is largely a consequence of the strong gravitational field of WD interiors; it modifies the composition and the opacity of the partially~-~degenerated outermost layers affecting its cooling. Another important effect is crystallization \cite{lambvanh75}. It is expected that Coulomb interactions inside carbon~-~oxygen WDs should be strong enough for forcing crystallization. This process should lead to a latent energy release that should be large enough to \textit{lengthen} WD cooling significantly. If a carbon~-~oxygen WD is massive enough it is expected to undergo crystallization at observable luminosity. Obviously, for the case of WDs made up of elements heavier than carbon and/or oxygen, crystallization should set in even earlier. Quite contrarily, for the case of He~WDs, the standard theory does not predict the occurrence of crystallization at luminosities reachable in real objects at the present age of the Universe.

He~WDs should arise as a consequence of binary evolution \cite{kippetal67}. A single star would need a time interval far in excess of the age of the Universe to evolve to such a state. Binary systems of close enough components lead to a mass exchange process in which the donor star, originally an intermediate mass object, losses most of its mass. Depending on the initial orbital separation and the masses of the components, it may lead to the formation of He~WDs on short times \cite{bendev05}.

He~WDs are expected to be the companions of some of the best observed binary systems containing a millisecond pulsar as are the cases of PSR~J0437-4715 \cite{vanstra01}, PSR~J1713+0747 \cite{spla05}, PSR~B1855+09 \cite{kasp94}, PSR~J1909-3744 \cite{jacob05}. Very recently \cite{stri09}, a population of He~WDs has been detected in the globular cluster NGC~6397. While most of the population of WDs is compatible with a carbon~-~oxygen rich interior, they found 24 serious candidates to be He~WDs.

Usually, the equation of state (EOS) of the He~WD is modeled as a degenerate Fermi gas of electrons with a free, uncondensed gas of He nuclei corrected by non~-~ideal effects (Coulomb, Thomas~-~Fermi, exchange, etc.). This is described in the appendix of Ref.~\cite{ab97}. Very recently, the possible occurrence of BEC in the deep interior of He~WDs has been studied. It has been proposed \cite{GabRos0809} that the ionic contribution to the specific heat of the condensate would be suppressed and that this would have observable consequences. Also the possibility that He~WDs would cool faster has been studied \cite{GabPir09} (for a review see Ref.~\cite{gabaros10}). The reason for this to be expected is very simple: at the interior of He~WDs, helium nuclei are light and close enough that their associated wave function can overlap that of neighboring nuclei. Consequently, the behavior of the gas of helium nuclei may depart from classical behavior allowing for the occurrence of quantum effects. 

Let us define the range of densities and temperatures expected to occur at He~WD interiors. These are dependent upon the mass of the object, however an absolute upper limit for the temperature $T$ is imposed by the conditions for helium burning: $T$ must be lower than $10^8$~K ($k_B T < 10$~KeV). Regarding the range of values for the density, for a given stellar mass value,  the upper limit is given by the central density of a zero temperature model as done by Hamada \& Salpeter \cite{hamsal61}. These values are given in Table~\ref{WDHE_hamsal}. It is important to remark that Nature does not provide a way to get higher densities with a helium-dominated composition avoiding helium burning. This is supported by  extensive numerical experiments performed by Iben \& Tutukov~\cite{ibentu85} who studied the evolution of close binary systems with components of intermediate masses. They found that the largest mass of He~WDs is $\approx$0.50~$M_{\odot}$. 

\begin{center} \begin{table}
\caption{\label{WDHE_hamsal} Hamada-Salpeter sequence for zero temperature helium white dwarfs. $M$ and $R$ are the stellar and radius, both in solar units, $\rho_{c}$ is the central density, $n_{c}$ is the central particle number density, and $T_{critical}$ is the critical temperature for the center of the star, defined by Eq.~(\ref{eq:tcrit}).}
\begin{tabular}{ccccc}
\hline
$M/M_{\odot}$  & 
$Log\; R/R_{\odot}$  & 
$Log\; \rho_{c}$ $[g/cm^3]$ & 
$n_{c}$ $[MeV^3]$ & 
$t_{critical}$ $[10^6~K]$ \\
\hline \hline 
0.50 & -1.864 & 6.292 & $2.340 \times 10^{-3}$ & 1.725 \\
0.45 & -1.842 & 6.164 & $1.742 \times 10^{-3}$ & 1.417 \\
0.40 & -1.817 & 6.029 & $1.277 \times 10^{-3}$ & 1.152 \\
0.35 & -1.792 & 5.882 & $9.103 \times 10^{-4}$ & 0.919 \\
0.30 & -1.764 & 5.721 & $6.283 \times 10^{-4}$ & 0.718 \\
0.25 & -1.734 & 5.539 & $4.132 \times 10^{-4}$ & 0.543 \\
0.20 & -1.699 & 5.325 & $2.524 \times 10^{-4}$ & 0.391 \\
0.15 & -1.657 & 5.061 & $1.374 \times 10^{-4}$ & 0.260 \\
0.10 & -1.602 & 4.704 & $6.042 \times 10^{-5}$ & 0.150 \\
\hline \hline \end{tabular} \end{table} \end{center}

Let us consider the range of temperatures for which we may expect the occurrence of quantum effects for the behavior of ions. Following \cite{gabaros10} the distance between $d$ ions should be of the order of the De Broglie wavelength $\lambda$. Then, the critical temperature $T_{critical}$ is given by

\begin{equation}
T_{critical}=  110.2\; \rho^{2/3}\; K \label{eq:tcrit} 
\end{equation}

Thus, $T_{critical}$ gives an estimation of the internal temperature at which the star may undergo some quantum mechanical effect in the ionic part of its composition, in particular a BEC. In the context of stellar interiors this temperature is very low. While in the literature it has been addressed the possible impact of such condensation on observable quantities \cite{GabPir09,gabaros10}, these analysis have been based on the simplified, analytical Mestel's treatment of WD cooling \cite{shap-teuk}.

In this paper we shall consider the cooling of He~WDs based on detailed numerical models in order to put some upper limit on the consequences of BEC on He~WDs evolution. In order to do so, we shall {\it not} perform a detailed treatment of the condensation but only mimic it by assuming that, for a given density, there exists a temperature $T_{cond}$ below which the ionic contributions to the energy, pressure and specific heats are strongly suppressed.

He~WDs may have or lack an outermost hydrogen layer. Such a hydrogen layer may be thick enough to undergo a strong nuclear activity at its bottom. This is found, for example as a consequence of binary stellar evolution \cite{bendev05}. For a given mass and luminosity, if He~WDs have such hydrogen layer, its interior is hotter than it would be without it. Consequently, the most favorable structure of condensation to occur is when the He~WDs has a helium~-~dominated composition up to its surface. Another relevant quantity to analyze is the amount of heavy elements (per unit of mass) $Z$  in the star. It is well known that, the higher $Z$ the higher the opacity of the plasma. Typical globular cluster have $Z$ values lower than that of the Sun ($Z_{\odot}= 0.020$).  Below we shall consider a set of models of He~WDs without any hydrogen layer, evolving them to very low luminosities to reach plausible conditions for BEC to occur.

The remainder of this paper is organized as follows: In Section~\ref{sec:model} we briefly describe the modifications we shall incorporate to the EOS of the stellar plasma together with the procedures to construct the stellar models. Section~\ref{sec:numerical} is devoted to present the numerical results of our study. Finally, in Section~\ref{sec:discuconclu} we discuss the meaning of our results related to the occurrence and detectability of BEC in
He~WDs and give some concluding remarks.

%------------------------------------------------------------------------------------------------------------------
\section{Modeling White Dwarfs} \label{sec:model} 

We shall employ the stellar evolution code described in \cite{bendev03} with updated physical ingredients. In order to mimic BEC, we shall consider it to occur at a temperature $T_{cond}= \eta T_{critical}$. $\eta$ has been incorporated in order to explore the effects of BEC for different values of the condensation temperature; we shall employ $\eta=$ 10, 3, 1, and 0. We shall consider an EOS as described in \cite{ab97} but for the ionic contributions to energy we incorporate a factor $F(T)$ defined as 

\begin{equation}\label{eq:fermi} 
F(T)= \bigg[ 1+\exp{\bigg(-\frac{T-T_{cond}}{\beta T_{cond}}\bigg)} \bigg ]^{-1}.
\end{equation} 

This function provides a modification to the ionic contributions to the EOS similar to those expected performing a detailed theory of BEC: for $T \gg T_{cond}$ we have the standard treatment without condensation whereas for $T \ll T_{cond}$ the ionic contributions to the EOS are negligible. $\beta$ has been introduced for numerical convenience in order to avoid too steep variations. While for the main part of this paper we have considered $\beta= 0.05$, we have also employed values of $\beta$ of 0.10 and 0.025. For these three  values of  $\beta$ the cooling curves to be presented below are hardly distinguishable, overlapping each other within the thickness of the lines in the plots. This is so even considering the fact that the shape of the function given in Eq.~(\ref{eq:fermi}) is very sensitive to the value of $\beta$. Other corrections relevant to the EOS (Coulomb, Thomas~-~Fermi, Exchange, etc. corrections) are assumed to be unaffected by BEC.

In order to construct the WD models we shall proceed as in \cite{ab97} and consider mass values of 0.30, 0.40, and 0.50~$M_{\odot}$ for two metallicity values: $Z=10^{-3}$ and $Z_{\odot}$. We start the computations assuming a polytropic structure that quickly relaxes to a WD configuration; doing so we avoid the complexity inherent to binary evolution. In any case, condensation is expected to occur at luminosities far lower than the initial ones; thus, the way we initialize our models has no impact on the main results of this work. 

%------------------------------------------------------------------------------------------------------------------
\section{Numerical Results} \label{sec:numerical}

\subsection{Cooling calculations}

\begin{figure} \epsfysize=500pt  \epsfbox{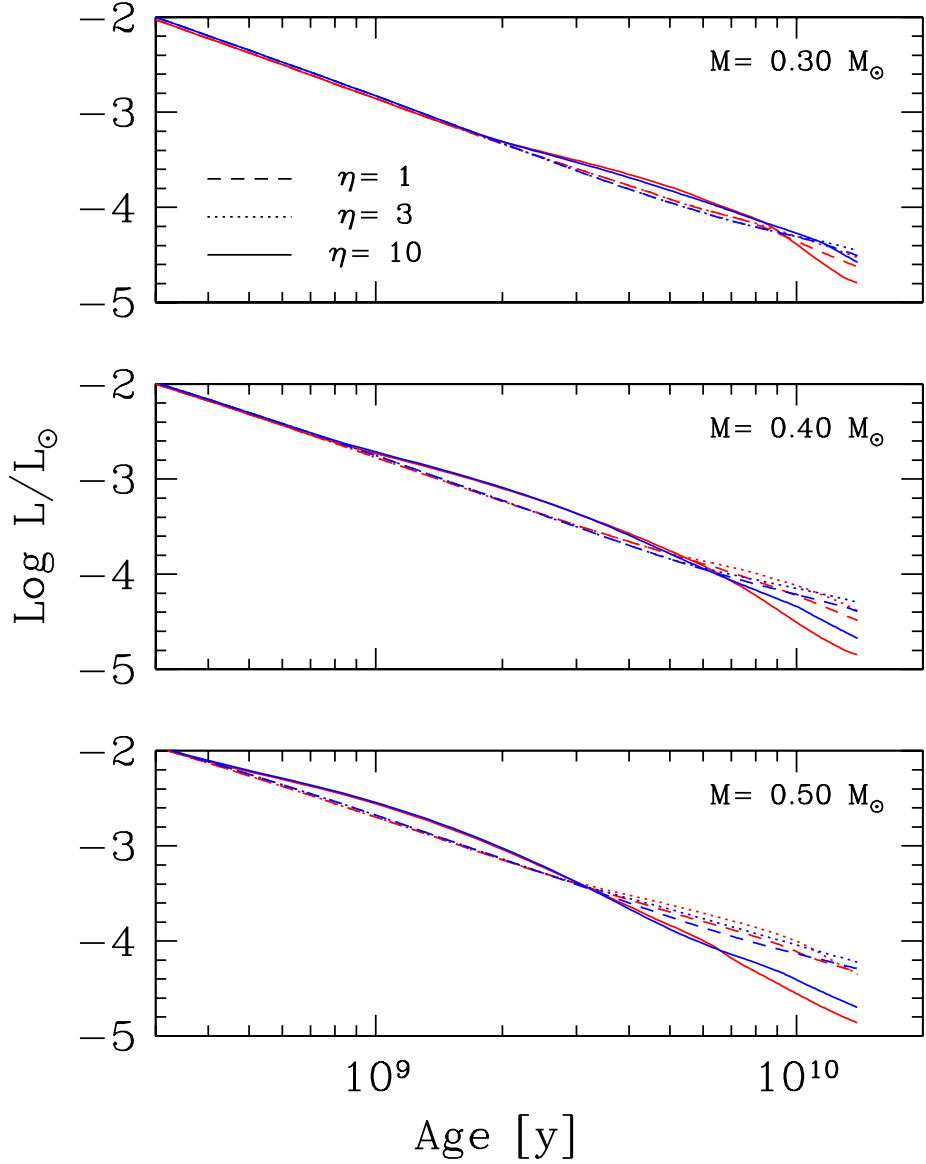} \caption{The luminosity of He~WDs as a function of the age for models of 0.3~$M_\odot$, 0.4~$M_\odot$ and  0.5~$M_\odot$ (upper, middle and lower panels respectively). Red (blue) lines correspond to a metallicity value of $Z=10^{-3}$ ($Z_{\odot}$). We have considered values of $\eta$= 1, 3, and 10 (dashed, dotted and solid lines respectively). Notice that the higher the mass of the WD, the effects due to BEC begin to be noticeable at higher luminosities and earlier ages. This is due to the fact that the central density of the models increases with the stellar mass. Notice that even if we consider large values of $\eta$, the effects due to the condensation are moderate.} \label{fig:cooling} \end{figure}

The cooling of our models is presented in Fig.~\ref{fig:cooling}. We have decided to present only the lowermost tail of the sequences because it is where the effects of BEC are noticeable. In this figure we did not included the results without BEC ($\eta=0$) because they are hardly distinguishable from those corresponding to $\eta=1$. The cooling curves are smooth and do not have any sudden change in its slope. 

It is well known that the specific heat of a helium nuclei gas cooling down to get condensation conditions, firstly grows above the classical value while later drops below it. The Mestel's analytic model of WD cooling predicts that the rate of change on the WD interior temperature is proportional to the inverse of the specific heat. Then we expect that, comparing to standard cooling without condensation, the first effect due to BEC is that the WD delays its cooling while later it is accelerated. This behavior is actually found in the detailed computations shown in Fig.~\ref{fig:cooling}: Comparing the cooling sequences for values of $\eta=$~3 and 10 with those corresponding to $\eta=1$ (very similar to those without BEC) the cooling sequences initially depart toward higher luminosities and then (if the WD gets old enough) drop below. As expected the effects due to BEC are more noticeable the larger the WD mass (the larger its central density) is. 

\begin{figure} \epsfysize=500pt  \epsfbox{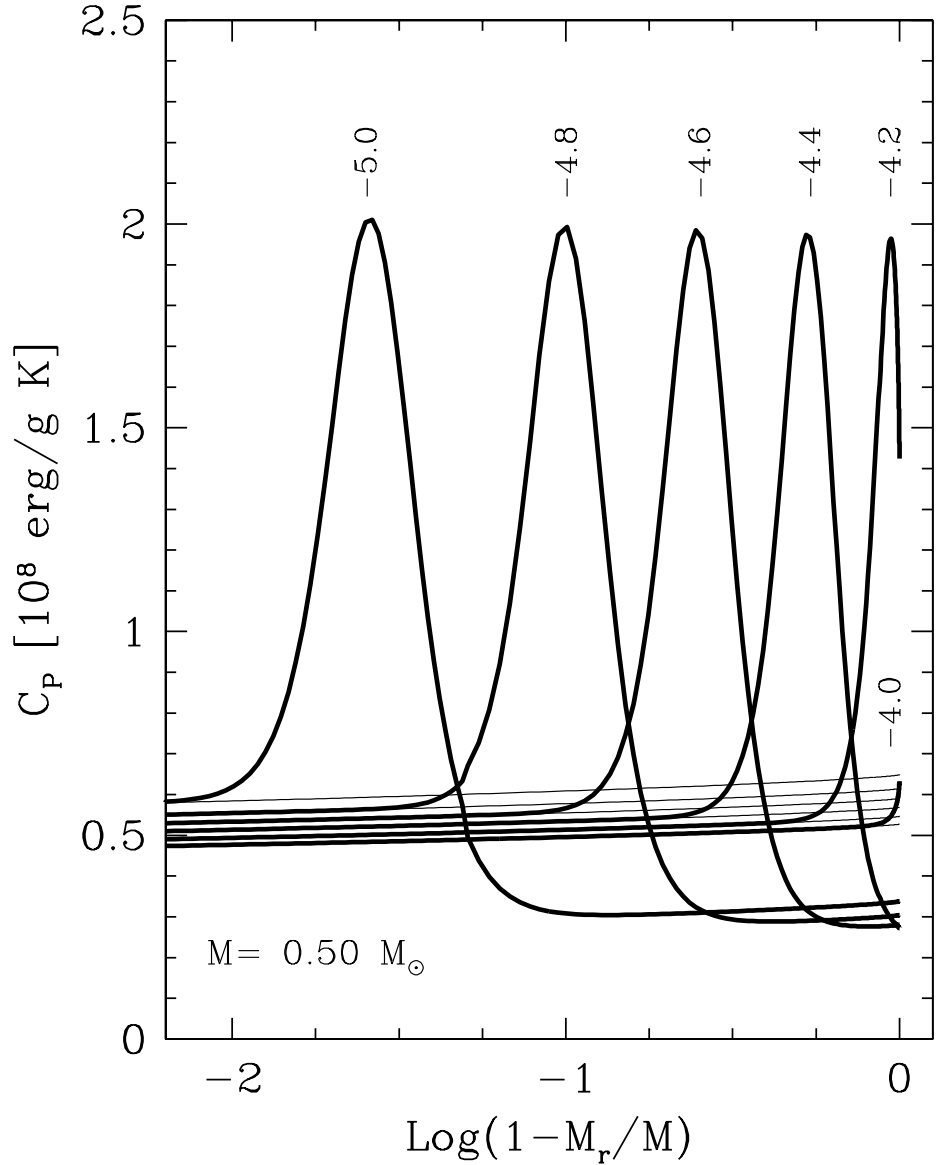} \caption{The evolution of the specific heat (at constant pressure) for a He~WD of 0.5~$M_\odot$. We assumed $\beta= 0.05$ and $\eta=10$. Thick lines represent the profiles corresponding to models including a suppression of the contributions due to helium nuclei devised to mimic BEC. For these models we included results for luminosities of $Log(L/L_\odot)=$ -4.0, -4.2, -4.4, -4.6, -4.8, and -5.0 which label the curves. Also, for the same luminosities, we included results corresponding models constructed assuming the standard treatment for the ionic contribution. These are denoted with thin solid lines and coincide with those corresponding to the same value of the luminosity for the outer layers of the star.} \label{fig:perfil_cp} \end{figure}

It is important to remark that these results, found mimicking BEC by introducing a factor given in Eq.~(\ref{eq:fermi}), should be largely independent of the details of BEC. This is so because of the dynamics of the process of condensation of the WD interior. The interior WD structure is dominated by a Fermi gas of strongly degenerate electrons; thus, as it cools down its density remains almost invariant. Since the onset of BEC at the stellar center, the fraction of condensed mass grows monotonously with time. It is important to remark that, while this process takes place, because its density (and also the condensation temperature) is much lower, the outermost envelope of the star  \textit{remains uncondensed}. The evolution of WDs is consequence of a balance of both of them: the effects due to BEC at the deep interior are averaged over the whole stellar structure. This is the reason why we do not find any sudden acceleration in the WD cooling, even considering the most massive He~WDs and $\eta=10$. This effect is described in Fig.~\ref{fig:perfil_cp}. There, we show the profile of the specific heat (at constant pressure) for the interior of a He~WD of 0.5~$M_\odot$ for very low luminosities of $Log(L/L_\odot)=$ -4.0, -4.2, -4.4, -4.6, -4.8, and -5.0. At the layers undergoing condensation we have an increase in $C_p$ above the standard values while for deeper layers $C_p$ is much lower than standard. However, it does not drop near to zero because of the contributions due to the degenerate Fermi gas of electrons and Coulomb interactions \cite{Han73}. The peak in the $C_p$ profile is due to the steep derivative of $F(T)$ for $T \approx T_{cond}$ (see Eq.~\ref{eq:fermi}). Obviously, a detailed treatment of BEC will provide another shape for such peak but again, this is not decisive for these cooling calculations because, as previously remarked, WD evolution is due to an average of the properties over its whole interior. 

\subsection{The Luminosity function}

The Luminosity function is the space density of a given type of star per unit interval of $Log(L/L_\odot)$ For the purposes of this paper we shall restrict ourselves to the simplest treatment of the luminosity function $\Phi(L/L_\odot)$. Following \cite{shap-teuk}, if in a given object (e.g., our Galaxy or a globular cluster) we may consider the space density and birth rate as uniform, then 

\begin{equation}
\Phi(L/L_\odot) \propto \bigg[ \frac{d}{dt} Log(L/L_\odot) \bigg ]^{-1}.
\end{equation}

\begin{figure} \epsfysize=500pt  \epsfbox{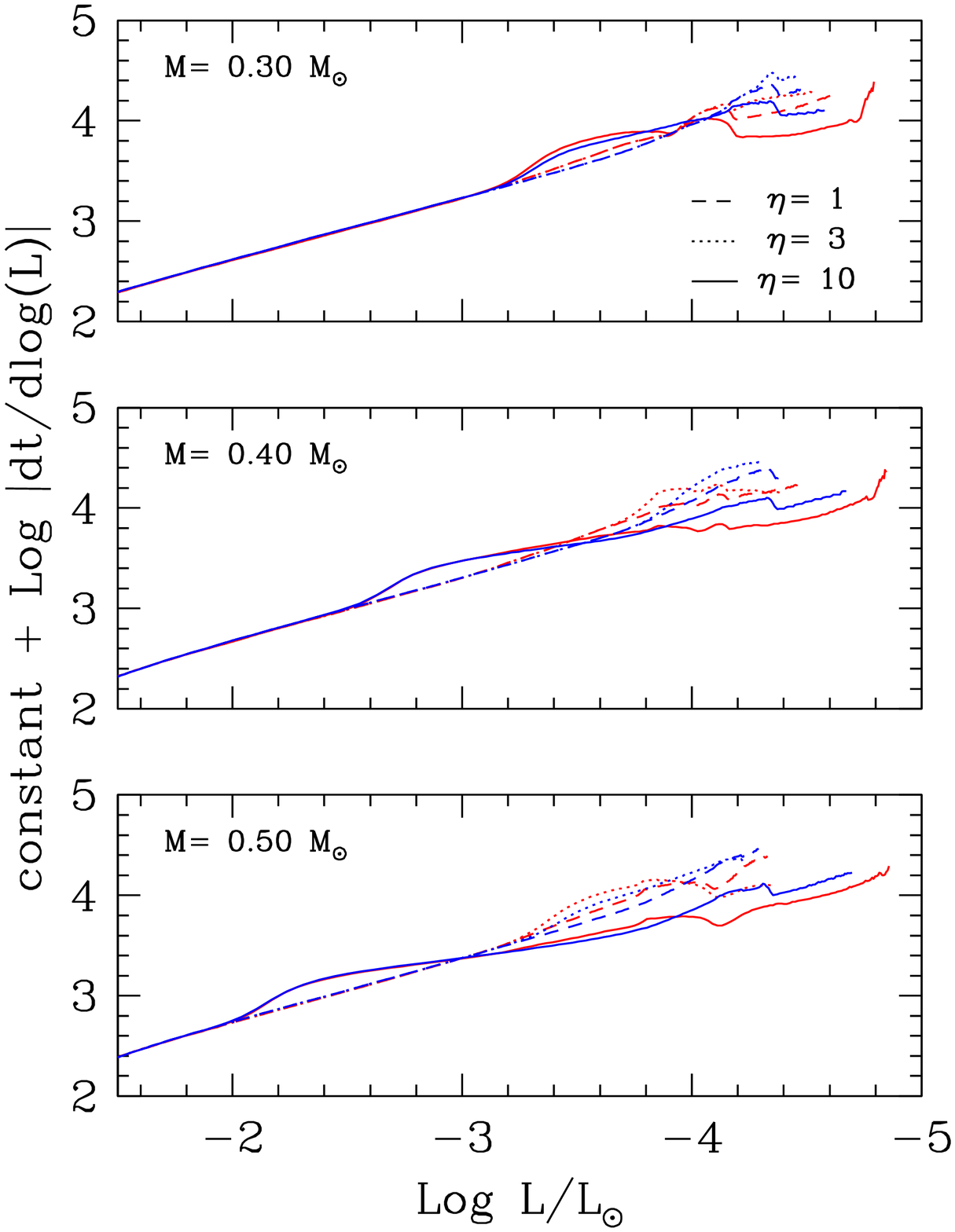} \caption{The luminosity function for He~WD models of 0.3~$M_\odot$, 0.4~$M_\odot$ and 0.5~$M_\odot$ (upper, middle and lower panels respectively). Red (blue) lines correspond to a metallicity value of $Z=10^{-3}$ ($Z_{\odot}$). We have considered values of $\eta$= 1, 3, and 10 (dashed, dotted and solid lines respectively). In view of the results presented in Fig.~\ref{fig:cooling} it is not surprising that the luminosity function undergoes moderate changes for the considered values of $\eta$. The dependence of the luminosity function with condensation is smaller than previous expectations (see Refs.~\cite{GabPir09,gabaros10}).} \label{fig:lumino} \end{figure}

In order to know $\Phi(L/L_\odot)$ we would need to know the stellar birth rate quantitatively. Fortunately this is not necessary for our purposes. Under these assumptions, we show in Fig.~\ref{fig:lumino} the logarithm of $\Phi(L/L_\odot)$. As remarked above, due to the smoothness of the cooling curves, the luminosity function corresponding to each WD mass value considered here does not undergo any steep drop due to BEC. The dependence of the luminosity function with condensation is far smaller than previous expectations (see Refs.~\cite{GabPir09,gabaros10}). This fact makes it {\it very difficult} to detect some observational signal of BEC in the cooling of He~WDs. Notice that we reached these conclusions considering the most massive He~WDs we expect to exist and high values for $\eta$.

%------------------------------------------------------------------------------------------------------------------
\section{Discussion and Conclusions}
\label{sec:discuconclu}

In this work we have computed a set of helium white dwarf (He~WD) models in order to explore the detectability of the effects due to a possible Bose~-~Einstein condensation (BEC) in their deep interior. To do so we have considered three values for the stellar mass (0.30, 0.40, and 0.50~$M_\odot$) and considered different possible values for the condensation temperature, whereas for the metallicity we assumed two values Z=$10^{-3}$ and 0.020. For the sake of simplicity, and because it provides the most favorable conditions for detecting BEC we neglected the possible existence of an outermost hydrogen layer. Here we do not perform a detailed theory of the behavior of a charged condensate as it has been done in Ref.~\cite{rosen10}. Because of the exploratory nature of this study, we decided to mimic the effects due to BEC in the EOS of the WD interior by introducing a factor represented by the function given in Eq.~(\ref{eq:fermi}) in order to suppress the ionic contribution when necessary. We should remark two advantages of this kind of procedure: First, the cooling of WDs is the result of an average of properties of matter over the entire stellar structure. This is a heat diffusion process that is well known to be fairly insensitive to short scale details. Thus, the cooling process is dependent on the fraction of mass that has a specific heat lower than that predicted by a classical statistics treatment. Because of these reasons, the procedure we employed is fairly adequate for our purposes. Second, it allows for an easy exploration of the space of parameters of the problem over a physically plausible space of parameters.

The absolute maximum density inside WDs with a helium dominated composition sets a very restrictive upper limit for the BEC to occur. The favorable environment is provided only for the case of the most massive He~WDs of, say, $\approx 0.50 M_\odot$ and with very low luminosities $Log(L/L_\odot) < -4.0$. Even in these conditions, while deep layers are condensed (and represent a monotonically growing mass fraction of the star), the outer layers {\it do not}. Moreover, the stellar surface has a direct connection with layers that never condensate. Thus, the cooling process is modified by the occurrence of BEC, but these signals are not dramatic. This is so for the two metallicity values here considered and even if we consider condensation temperatures very high (as compared to that predicted by Eq.~\ref{eq:tcrit}). These facts lead us to conclude that it seems very difficult to find observable signals of BEC. 

As stated in the Introduction (\S~\ref{sec:intro}), a population of 24 candidates to be He~WDs has been detected \cite{stri09} in the globular cluster NGC~6397. Recently, it has been proposed \cite{GabPir09,gabaros10} that such population may show some signal of BEC. Let us analyze this possibility in the frame of the results presented above. Six of the 24 candidates proposed in \cite{stri09} to be He~WDs, representative of the whole population in NGC~6397, have been previously detected \cite{cool98,tayl01}. In \cite{cool98} it has been found that these objects are He~WDs with masses of the order of $M \sim 0.25~M_{\odot}$. In \cite{edm99} it has been reported the spectroscopy of one of these objects (C2, see Table~\ref{table:observations}), showing that it has $log \, g \simeq 6.25 \pm 0.1$, while Ref.~\cite{cool98} provides their absolute visual magnitude. Adopting the same modulus of distance (as done in \cite{cool98}) of $(m - M)_0 = 11.71$ for NGC~6397  for the subset of candidates studied in Ref.~\cite{tayl01},  and employing bolometric correction for He~WD model with $T_{eff}=8000$~K and $log \, g = 6.25$ from  Ref.~\cite{rhor01}, we compute the values of $Log (L / L_{\odot})$. The results are given in Table~\ref{table:observations}. We found a minimum value of $Log (L / L_{\odot}) \sim -2.5$.

\begin{table}
\caption{\label{table:observations} Six  candidates for He~WDs in the globular cluster NGC~6397. They are identified (ID) as C1, C2 and C3 corresponding to Ref.~\cite{cool98} and T1, T2 and T3 to Ref.~\cite{tayl01}. See text for details.}
\begin{center} \begin{tabular}{lrc}
\hline\hline 
ID  & $M_V$  & $Log (L/L_{\odot})$ \\
\hline 
C1 &  6.79 & -0.731 \\
C2 &  7.84 & -1.151 \\
C3 &  8.06 & -1.239 \\
T1 & 11.05 & -2.435 \\
T2 & 11.05 & -2.435 \\
T3 & 11.34 & -2.551 \\
\hline \hline \end{tabular}  \end{center} \end{table} 

Comparing the results presented in Section~\ref{sec:numerical} with the data presented in Table~\ref{table:observations} we find that the He~WDs population in NGC~6397 did not undergo BEC yet: their masses are too low and are too bright. Thus, unfortunately, NGC~6397 does not provide an adequate laboratory to look for signals of the charged condensate.

We would like to thank our anonymous referee for his/her constructive report that helped us in improving the original version of this paper.
 
%------------------------------------------------------------------------------------------------------------------

\end{document}